\documentstyle[12pt]{article}
\newcommand{\be}[1]{\begin{equation} \label{(#1)}}
\newcommand{\ee}{\end{equation}}
\newcommand{\ba}[1]{\begin{eqnarray} \label{(#1)}}
\newcommand{\ea}{\end{eqnarray}}
\newcommand{\nn}{\nonumber}
\newcommand{\rf}[1]{(\ref{(#1)})}

\def\rp{$R_p \hspace{-1em}/\;\:$}
\def\rpm{R_p \hspace{-0.8em}/\;\:}
\def\rpt{$R_p \hspace{-0.85em}/\ \ $}

\def\pmb#1{\setbox0=\hbox{#1}%
  \kern-.015em\copy0\kern-\wd0
  \kern.03em\copy0\kern-\wd0
  \kern-.015em\raise.0233em\box0 }
\def \znbb {0\nu\beta\beta}

\def\bfr{\pmb{${r}$}}
\def\bfsgm{\pmb{${\sigma}$}}
\def\sir{({ \bfsgm_{i}^{~}} \cdot {\hat{\bfr}_{ij}^{~}} )}
\def\sjr{({ \bfsgm_{j}^{~}} \cdot {\hat{\bfr}_{ij}^{~}} )}
\def\si{{ \bfsgm_{i}^{~}}}
\def\sj{{ \bfsgm_{j}^{~}} }
\begin{document}
\begin{center}
{\bf On  Prospects for Exploration of Supersymmetry
        in Double Beta Decay Experiments.}

\bigskip

V.A. Bednyakov,  V.B. Brudanin, S.G. Kovalenko \\
and\\
 Ts.D. Vylov,\\
{\it Joint Institute for Nuclear Research, Dubna, Russia}\\[5mm]
\end{center}

\bigskip

\begin{abstract}
	We analyze constraints on the parameters of the $R_p$ violating
	supersymmetry (\rp SUSY) which can be extracted from non-observation
        of the neutrinoless nuclear double beta decay ($\znbb$) at a given
	half-life lower bound.
	Our analysis covers a large class of phenomenologically viable
	\mbox{\rp SUSY} models.
	We introduce special characteristics:
        the SUSY sensitivity of a $\beta\beta$-decaying isotope
	and the SUSY reach of a $\znbb$-experiment.
        The former provides a physical criterion for a selection of the
        most promising isotopes for SUSY searches and  the latter
	gives a measure of success for a $\znbb$-experiment in
        exploring the \rp SUSY parameter space.
	On this basis  we discuss prospects for exploration of supersymmetry
	in various $\znbb$-experiments.
\end{abstract}

\newpage
        Observation of the neutrinoless double beta ($\znbb$) decay
$ {}^{Z}_A Y\rightarrow {}^{Z+2}_{~~~A} Y + 2 e^-$ of some nucleus ${}^{Z}_A Y$
        would be an unambiguous signal of the physics beyond the
	standard model (SM) of electro-weak interactions
\cite{hax84},~\cite{doi85}.
	This process is forbidden within the SM since it violates lepton
	number (L) conservation.

        During a long time the $\znbb$-decay has been attracting great
        theoretical and experimental efforts.
        A number of experimental collaborations are involved in searching for
        this exotic phenomenon (see for instance
\cite{rev_exp}).
        Unfortunately, there is not yet any evidence for the $\znbb$-decay.
        Nevertheless, impressive progress has been achieved in
        establishing a lower bound on the half-life of various isotopes
	and further advance in this direction is expected in the near future.

        The question is:
	what sort of new information on the physics beyond the SM
        we are provided with this experimental progress.

	It was a common practice to answer this question in terms of
	the Majorana neutrino mass.
        This implies that the $\znbb$-decay proceeds via the conventional
	mechanism based on Majorana neutrino exchange
        between the decaying neutrons, as presented in Fig.~1a.
	The measure of experimental success in this case is a reduction of
	the upper bound on the effective neutrino mass
	$\langle m_{\nu} \rangle$.
	The best result to date is \mbox{$\langle m_{\nu} \rangle
	 \leq 0.65$~eV (90\% C.L.)}
\cite{hdmo95}.
	Other experiments are also nearly penetrating the sub-eV
        neutrino mass range.

	Quite recently it has been realized that the conventional
        neutrino exchange mechanism is not the only possible one.
        Modern particle physics offers a mechanism based on the
	supersymmetric (SUSY) interactions
\cite{Mohapatra}-\cite{MHKK}.
	In a certain sense this mechanism is more interesting than
        the conventional one since it allows the $\beta\beta$-decay
        experiments in the exciting field of supersymmetry.
        SUSY interactions, inducing the $\znbb$-decay, can contribute
	to many other processes.
	Therefore, one can compare information about these
	interactions obtained from different
	experiments including $\znbb$-decay experiments.
	It has been shown in
\cite{HKK2} that $\znbb$-decay experiments are more sensitive to
	certain SUSY manifestations than the other running
	and forthcoming accelerator and non-accelerator experiments.

        In this note we give a quite general parametrization
        of the SUSY effect in the $\znbb$-decay and formulate criterion for
        the success of a $\znbb$-experiment in terms of
        an upper bound on the effective SUSY parameter.
        The latter is a certain combination of the fundamental
        SUSY parameters.
	We introduce the sensitivity of an isotope to the SUSY manifestation
	and the SUSY reach of a given $\znbb$-experiment characterizing
        the depth of its penetrating to the SUSY realm.
        The latter is represented by the unexplored part of the SUSY
	model parameter space.
        This provides us with a convenient basis for estimating prospects of
	the exploration of the SUSY in various $\znbb$-experiments.

	We start with a short theoretical introduction.
        The $\znbb$-decay is allowed within a special class of the SUSY
	models admitting B-L violation
	(B and L are baryon and lepton quantum numbers).
	These are presently popular SUSY models with R-parity violation
	($R_p = (-1)^{3B+L+2S}$, where $S$ is the spin).
	The \rpt terms can be introduced explicitly into the
	superpotential as
\ba{superpotential}
W_{\rpm} &=& \lambda_{ijk}L_i L_j {\bar E}_k
           + \lambda'_{ijk}L_i Q_j {\bar D}_k
           + \lambda''_{ijk}{\bar U}_i {\bar D}_j {\bar D}_k.
\ea
        The indices $i,j, k$ stand for generations.
	$L$, $Q$ denote lepton quark doublet superfields, while
	${\bar E}, \ {\bar U},\  {\bar D}$ correspond to
	lepton and {\em up}, {\em down} quark singlet  superfields.
	The first two terms in
\rf{superpotential} lead to lepton number violation,
        while the last one violates the baryon number.
        For the $\znbb$-decay only the $\lambda$ and $\lambda'$ type
	couplings are of relevance.
	The presence of lepton number violating interactions in the
	\rp SUSY model allows one to construct the SUSY mechanism
        of the $\znbb$-decay
\cite{Mohapatra}-\cite{MHKK}.

	A complete analysis of this mechanism within the \rpt minimal
	supersymmetric standard model (\rp MSSM) was carried out in
\cite{HKK2}.
        On rather general grounds it was also shown
\cite{HKK2} that the dominant contribution to this mechanism within
	a phenomenologically viable \mbox{\rp SUSY} model comes from the
	gluino $\tilde g$ exchange diagram presented in Fig.~1b.
	This contribution does not depend on specific details of a
	\mbox{\rp SUSY} model such as the neutralino content, the mixing
	coefficients, etc.
        Thus, for any such model we can write down a dominant
	term of the effective quark-electron Lagrangian
        in the same form as for the \rp MSSM derived in
\cite{HKK1}
\ba{Leta}
{\cal L}_{eff}(x)\ &=&\
\frac{G_F^2}{2 m_p}\cdot  \eta_{SUSY}
\left[J_{P}J_{P} + J_S J_S - \frac{1}{4} J_T^{\mu\nu} J_{T \mu\nu}\right]
(\bar e (1 + \gamma_5) e^{\bf c}) \\ \nonumber
&+& \mbox{subdominant terms}.
\ea
	The quark currents are defined as
$
J_{P} =   \bar u^{\alpha} \gamma_5 d_{\alpha}, \
J_{S} =   \bar u^{\alpha} d_{\alpha}, \
J_T^{\mu \nu} = \bar u^{\alpha} \sigma^{\mu \nu}(1 + \gamma_5) d_{\alpha}
$.
       The effective parameter $\eta_{SUSY}$
        accumulates certain fundamental parameters of the 
	\mbox{\rp SUSY} model.
        In the particular case of the \rp MSSM it has the form
\ba{eta1}
\eta_{\tilde g} &=& \frac{2 \pi \alpha_s}{9}
\frac{\lambda^{'2}_{111}}{G_F^2 m_{\tilde d_R}^4}
\frac{m_p}{m_{\tilde g}}\left[
1 + \left(\frac{m_{\tilde d_R}}{m_{\tilde u_L}}\right)^4\right].
\ea
        Here  $m_{\tilde g}$ is the gluino mass and
        $m_{\tilde u_L, \tilde d_L}$ are masses of the superpartners  of
        $u_L$ and $d_R$ quarks; $\alpha_s$ is the strong coupling constant.
	Starting from the Lagrangian
\rf{Leta} describing the basic quark-lepton transition
        $d+d\rightarrow u+u+2e^-$, which triggers the $\znbb$-decay,
	one can derive the following half-life formula
\cite{HKK1}
\be{tht12}
\big[ T_{1/2}(\znbb) \big]^{-1} =
 G_{01} \left(\frac{m_A}{m_p}\right)^4 \eta_{SUSY}^2
\left|{\cal M}_{\tilde q}\right|^2.
\ee
        The factor in brackets implies the normalization of
        ${\cal M}_{\tilde q}$ as in
\cite{HKK2}.
	$m_A =0.85$~GeV is the momentum scale of the nucleon dipole
	form factor $F(q^2) = (1 + q^2/m_A^2)^{-2}$ used in
	calculation of the nuclear matrix element ${\cal M}_{\tilde q}$.
  	The leptonic phase space integral $G_{01}$ is defined as
\be{G_01}
 G_{01} = \frac{(G_F \cos\Theta_C f_A/f_V)^4 m_e^9}
	       {64 \pi^5 \hbar (m_e R_0)^2 \ln 2}
	       \int d\Omega_1 d\Omega_2 b_{01}
\ee
	and tabulated in
\cite{doi85} for various nuclei. Values of $G_{01}$ for nuclei analyzed
        in the present letter are given in the Table.
	Here, $R_0$ is the nuclear radius and $f_A = 1.261$, $f_V=1$.
	The kinematical factor $b_{01}$ accounts for the Coulomb distortion
        of the electron waves
\cite{doi85}.

	The nuclear matrix element ${\cal M}_{\tilde q}$ for the SUSY
        contribution to the $\znbb$-decay can be written in the form
        independent of the nuclear wave function
\cite{HKK1}
\ba{matr}
{\cal M}_{\tilde q} &=& {\frac{m_p}{ m_e}}
\big< 0^+_f || \sum_{i \ne j} \tau_{+}^{(i)} \tau_{+}^{(j)}
                     \left(\frac{R_0}{r_{ij}}\right)
\left[ \alpha_V^{(0)} F_{N}(x_{A}) +  \alpha_V^{(1)} F_{4}(x_{A})
+ \right.\\ \nn
&+& \left.\si \cdot \sj \left(\alpha_A^{(0)} F_{N}(x_{A}) +
\alpha_A^{(1)} F_{4}(x_{A})\right) +
\alpha_T\big\{ 3 \sir \sjr  - \right. \\ \nn
&-& \left. \si \cdot \sj \big\} F_{5}(x_{A})
\right] || 0^+_i \big>.
\ea
	The following notations are used
$$
{\bfr}_{ij} = ({\overrightarrow r}_i - {\overrightarrow r}_j ),\ \ \
r_{ij} = |{\bfr}_{ij}|, \ \ \ {\hat{\bfr}_{ij}^{~}}= {\bfr}_{ij}/r_{ij},\ \ \
x_A = m_A r_{ij},
$$
        where ${\overrightarrow r}_i$ is the coordinate of the {\it i}th
        nucleon.
	The nucleon structure parameters
	$\alpha_V^{(0)} = 0.15$, $\alpha_V^{(1)} = 7.5$,
	$\alpha_A^{(0)} = -1.2$, $\alpha_A^{(1)} = 0.28$,\
	$\alpha_T = 1.3$ were calculated in
\cite{HKK1}.

	Three different structure functions $F_i$ are given by the
	integrals over the momentum ${\bf q}$ transferred between
        two decaying nucleons and have the form
\be{Fnn}
F_N(x)=\frac{xe^{-x}}{48}(3 + 3 x + x^2),\ \
F_4(x)=\frac{xe^{-x}}{48}(3 + 3 x - x^2),\ \
F_5(x)  = \frac{x^3e^{-x}}{48}.
\ee
        Expression
\rf{matr} allows numerical calculation of the matrix element
	${\cal M}_{\tilde q}$ within any nuclear structure model.
	So far such calculations have been carried out only within
        the pn-QRPA (proton-neutron Quasiparticle Random Phase
	Approximation) in
\cite{HKK2}.
        It is well-motivated and reliable approach
\cite{mut89} successfully describing the $2\nu\beta\beta$-decay recently
	observed in several experiments.
	Numerical values of ${\cal M}_{\tilde q}$ for several experimentally
	most interesting isotopes are given in the Table.
	We point out that this matrix element drastically differs in structure
	from the well known $\znbb$ nuclear matrix elements of
	the Majorana neutrino exchange mechanism.

        Non-observation of the $\znbb$-decay at the half-life level
        $T_{1/2}^{exp}$ leads to the following constraint on the \rp SUSY
        parameter $\eta_{SUSY}$
\be{con}
	T_{1/2}^{th}\geq T_{1/2}^{exp} \longrightarrow\
        \eta_{SUSY} \leq \frac{10^{-7}}{\zeta(Y)}
	\sqrt{\frac{10^{24} \mbox{years}} {T_{1/2}^{exp}}} =
        10^{-7}\cdot \epsilon(Y,exp)^{-1},
\ee
	where $T_{1/2}^{th}$ is the theoretical expression given in
\rf{tht12}.
        We have introduced the following characteristics of a
        $\beta\beta$-decaying isotope $Y$ and a particular
	$\znbb$-experiment giving a quantitative basis for assessing
        its ability to search for a SUSY signal.
        These are the SUSY sensitivity of an isotope $Y$
\be{sss}
\zeta(Y) = 10^{5}{\cal M}_{\tilde q}\sqrt{G_{01}}
\ee
        and the SUSY reach of the $\znbb$-experiment with the isotope $Y$
\be{reach}
\epsilon(Y,exp) = \zeta(Y)
	\sqrt{\frac{T^{exp}_{1/2}}{10^{24} \mbox{years}}}.
\ee
        The former is an intrinsic characteristic of an isotope
        $Y$ depending only on the matrix element and the phase space
        factor.
        The latter characterizes the experimental set-up as a whole.
        $T^{exp}_{1/2}$ is the lower half-life bound reached with this set-up.
	This is a time dependent characteristic for improving experiments.

	The large numerical values of the SUSY sensitivity $\zeta$ in
\rf{sss} correspond to those isotopes within the group of
	$\beta\beta$-decaying nuclei which are most promising
	candidates for searching for SUSY in $\znbb$-decay experiments.
	Using general formulas in
\rf{matr}-\rf{Fnn} one can calculate $\zeta$ in the framework of any
	nuclear structure model.
        Its numerical values calculated in the pn-QRPA are
        presented in the Table.
        As is seen, the most sensitive isotope is $^{150}$Nd, then
        follows $^{100}$Mo.
        Unfortunately, we are not able to make any conclusion concerning
        $^{48}$Ca, which is also an experimentally interesting isotope.
        This is because the pn-QRPA does not provide reliable results for
	this light nucleus.
	It would be important to reconstruct a whole picture showing
	potential abilities of all $\beta\beta$-decaying nuclei
	from the point of view of their sensitivity to the SUSY signal.
	Therefore, calculation of the SUSY nuclear matrix element
        ${\cal M}_{\tilde q}$ for $^{48}$Ca within a certain nuclear
	model is demanded.
        It could be done, for instance, on the basis of the nuclear shell.
        This approach  allows
        exact nuclear structure calculations for $^{48}$Ca
\cite{Haxton},~\cite{Retamosa}.

        Of course, the SUSY sensitivity $\zeta$ cannot be the only criterion
        for selecting an isotope for the $\znbb$-experiment. 
	Other microscopic and
        macroscopic properties of the isotope are also important for building
        a $\znbb$-detector.

        The SUSY reach parameter $\epsilon(Y, exp)$
	of the $\znbb$-experiment takes into account both the SUSY
        sensitivity of the isotope $Y$ and specific experimental
        conditions determining the reach in the half-life limit.
	The current and the near future experimental situation in terms of
	the accessible half-life and the SUSY reach parameter $\epsilon$
	is presented
in Fig.~2a,b.
        For completeness,
in Fig.~2c we also present a situation with the experimental reach in the
	inverse neutrino mass.
	Fig.~2c is taken from
\cite{hdmo95}.
        Figs. 2 show the running $\beta\beta$-experiments:
        the Heidelberg-Moscow (H-M) $^{76}$Ge experiment
\cite{hdmo95},
	the NEMO $^{100}$Mo experiment
\cite{nemo94}, and other $\znbb$-experiments that also crossed
        the half-life barrier of $10^{21}$ years with the isotopes
	$^{48}$Ca
\cite{you95}; $^{82}$Se
\cite{ell92}; $^{100}$Mo
\cite{als93}; $^{116}$Cd
\cite{dan95}; $^{130}$Te
\cite{ale94}; $^{136}$Xe
\cite{vui93}; $^{150}$Nd
\cite{moe94}.
        We also included in Fig. 2  the forthcoming or partly operating
        experiments: the Gottard $^{136}$Xe TPC experiment
\cite{jorg94}, the $^{130}$Te cryogenic experiment
\cite{ale94}, the ELEGANT experiment with 64 $g$ of $^{48}$Ca
\cite{elegant95}, the experiment with 1.6 $kg$ of $^{150}$Nd
	using an improved UCI TPC
\cite{moe94}.

In Fig.~2a two broken lines with black squares and triangles
	correspond to the two fixed values of the SUSY reach
	$\epsilon({}^{76}\mbox{Ge, H-M})^{\mbox{present}} = 5.3$ (lower
	line) and
	$\epsilon({}^{100}\mbox{Mo, NEMO})^{\mbox{future}} = 22.2$
	(upper line).
        Experiments reaching these lines provide the same constraints
        on $\eta_{SUSY}$ in
\rf{con} as the Heidelberg-Moscow experiment at present (lower line)
        and as expected from the NEMO experiment in the 
	near future (upper line) respectively.

        These lines are interrupted on the left from the $^{76}$Ge
        since we have no value of the SUSY  nuclear matrix element
        ${\cal M}_{\tilde q}$ for $^{48}$Ca.

	It is seen from
Fig.~2 that the Heidelberg-Moscow $^{76}$Ge detector
	is the best $m_{\nu}$ explorer at present and up to the year 2000.
	This experiment is also the best SUSY explorer at present.
        However, in the near future the NEMO experiment is able to take over
        the leadership in exploration of SUSY as follows from
Fig.~2a.

        In conclusion, we stress that from the physical point of view
        the lower half-life bound itself is not a satisfactory
        characteristic of the $\znbb$-decay experiment.
        An objective characteristic
        should be related to the fundamental parameters of the underlying
        physics. In this note we give the parameterization of
        the SUSY contribution to the $\znbb$-decay in terms of one
        effective parameter $\eta_{SUSY}$. The parameterization is valid for
        a wide class of \rp SUSY models and can be used for
        the quantitative presentation of the SUSY effect in the $\znbb$-decay
        experimental data. Together with the effective
        Majorana neutrino mass 
	$\langle m_{\nu} \rangle$ the parameter $\eta_{SUSY}$
        is an important physical characteristic of
        the $\znbb$-decay experiment.

        Our results for the SUSY sensitivities $\zeta$
        of various $\beta\beta$-decaying isotopes and
        for the SUSY reach $\epsilon$ of $\beta\beta$-decay detectors
        based on  the $\eta_{SUSY}$ parameterization
        can be used for planning the future experiments.

\bigskip

{\bf Acknowledgments}

\smallskip
        We thank M. Hirsch and H.V. Klapdor-Kleingrothaus for
        fruitful discussions.
        This work was supported in part by Grant 215 NUCLEON from
        the Ministry of
        Science and Technological Policy of the Russian Federation.


\begin{table}[t]
{Table: The SUSY sensitivity $\zeta$, phase space factor $G_{01}$ and
nuclear matrix element for the SUSY mechanism of the $\znbb$-decay
calculated within the pn-QRPA for several experimentally interesting isotopes.
$\zeta$ defines the constraint on the SUSY parameter
$\eta_{SUSY} \leq \frac{1}{\zeta\cdot 10^{7}}
\left( \frac{10^{24}}{T_{1/2}^{\znbb}}\right)^{1/2}$.}\\[3mm]
\begin{center}
\begin{tabular}{|c|c|c|c|}\hline &&&\\
Isotope &Nuclear Matrix&Phase Space& SUSY\\
&Element &Factor &Sensitivity \\
$^{A}$Y &  ${\cal M}_{\tilde q}$ & $G_{01}\cdot 10^{14}$   & $\zeta$\\
&&&\\ \hline &&&\\
$^{150}$Nd  & 416 & 20.808   & 19     \\ &&&\\ \hline &&&\\
$^{100}$Mo  & 328 & 4.56316   & 7     \\ &&&\\ \hline &&&\\
$^{130}$Te  & 262 & 4.41596  & 5.5    \\ &&&\\ \hline &&&\\
$^{82}$Se   & 253 & 2.80583   & 4.24  \\ &&&\\ \hline &&&\\
$^{116}$Cd  & 190 & 4.92614  & 4.2    \\ &&&\\ \hline &&&\\
$^{136}$Xe  & 143 & 4.71196  & 3.11   \\ &&&\\ \hline &&&\\
$^{76}$Ge   & 283 & 0.635941  & 2.26  \\ &&&\\ \hline &&&\\
$^{128}$Te  & 298 & 0.181888 & 1.27   \\ &&&\\ \hline
\end{tabular}
\end{center}
\end{table}

\newpage

{\large\bf Figure Captions}\\

\begin{itemize}

\item[Fig.1]    (a) The conventional massive Majorana neutrino
                    exchange mechanism and
                (b) the supersymmetric mechanism
                    (the dominant contribution)
                     of the neutrinoless double beta decay.

\item[Fig.2]    1995 experimental situation (gray bars)
                and expectations up to 2000 (open bars).
                (a) half-life $T_{1/2}$;
		two broken lines correspond to the two fixed values
		of the SUSY reach
		$\epsilon({}^{76}\mbox{Ge, H-M})^{\mbox{present}} = 5.3$
                (lower line) and
		$\epsilon({}^{100}\mbox{Mo, NEMO})^{\mbox{future}} = 22.2$
		(upper line);
		experiments reaching these lines provide the same
		constraints as the Heidelberg-Moscow experiment now
		(lower line)
                or as expected from the NEMO experiment in the near
                future (upper line).
                (b) the SUSY reach parameter $\epsilon$ (see \rf{reach})
                and
		(c) the inverse neutrino mass reach $1/m_{\nu}$ (taken from
                \cite{hdmo95}).
\end{itemize}
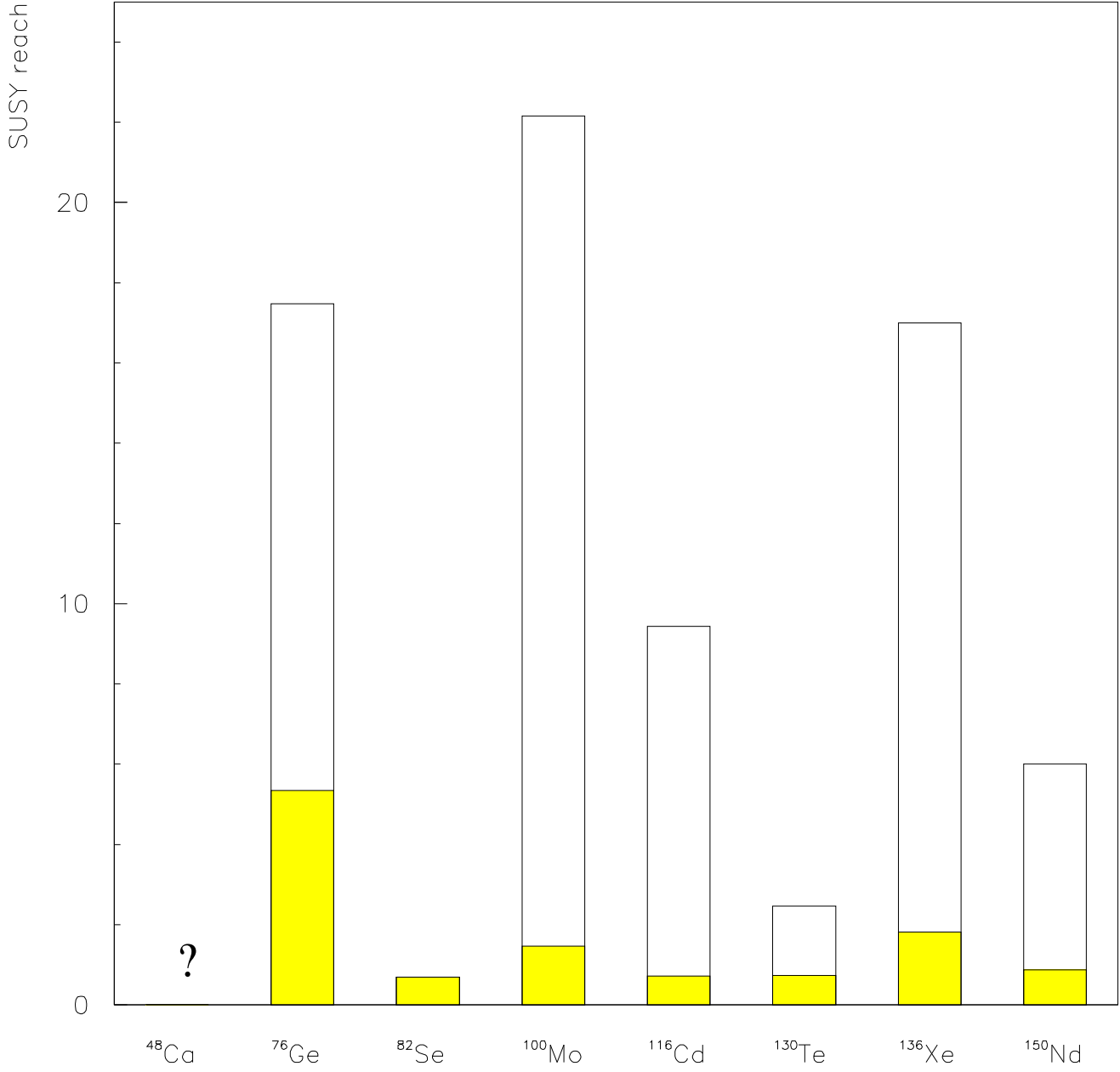
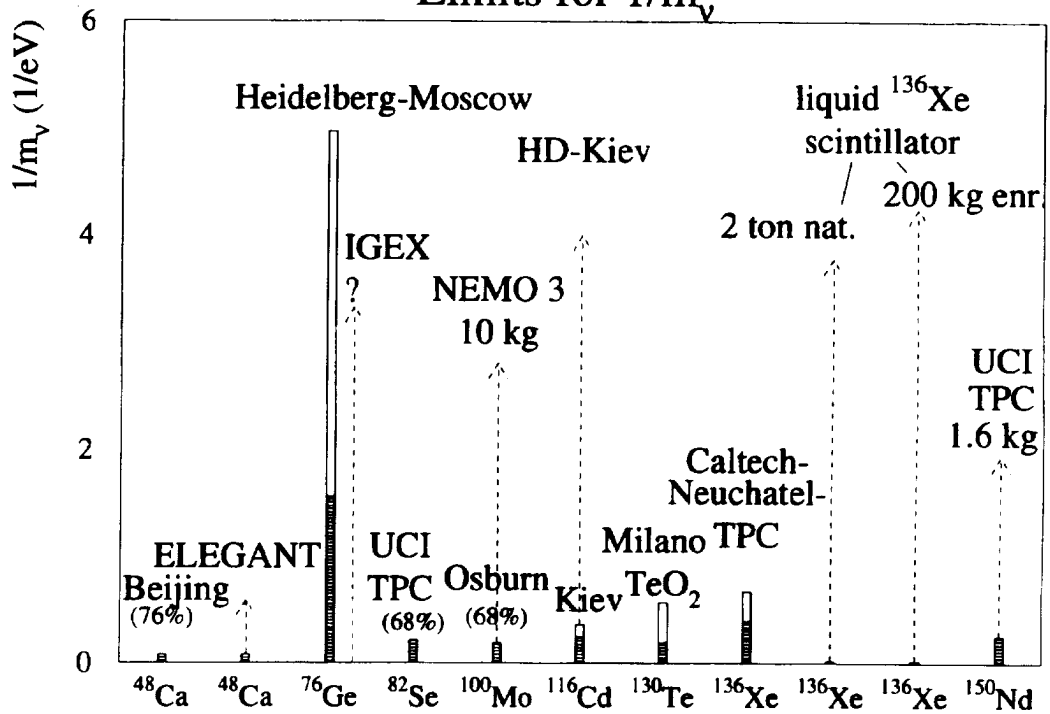
\end{document}